\documentclass[11pt,a4paper]{article}
\usepackage[utf8x]{inputenc}
\usepackage{amsfonts}
\usepackage{amsmath}
\usepackage{amssymb}
\usepackage{dsfont}
\usepackage{cite}
\usepackage{geometry}
\usepackage{enumerate}

\newcommand{\rme}{\mathrm{e}}
\newcommand{\rmi}{\mathrm{i}}
\newcommand{\rmd}{\mathrm{d}}
\renewcommand{\qquad}{\hspace*{25pt}}

\newcommand{\tr}{\mathop{\mathrm{tr}}\nolimits}

\def\pt(#1){({\it #1\/})}

\newcommand{\JPA}{{\it J. Phys. A: Math. Gen.} }

\newcommand{\QSO}{{\em Quantum Semiclass. Opt.} }

\newcommand{\AP}{{\it Ann. Phys., Lpz.} }

\newcommand{\JCP}{{\it J. Chem. Phys.} }

\newcommand{\JMP}{{\it J. Math. Phys.} }

\newcommand{\PL}{{\it Phys. Lett.} }
\newcommand{\PR}{{\it Phys. Rev.} }
\newcommand{\PRL}{{\it Phys. Rev. Lett.} }

\newcommand{\nonum}{\par\item[]} 

\title{\bf Non-Markovian quantum trajectories, instruments and
time-continuous
measurements}

\author{\\{\large \bf Sven
Kr\"onke\footnote{Sven.Kroenke@physnet.uni-hamburg.de}}\\{\small \it Zentrum
f\"ur optische Quantentechnologien, Universit\"at Hamburg,}\\{\small \it D-22761
Hamburg, Germany}\\\\
{\large\bf  Walter T. Strunz\footnote{Walter.Strunz@tu-dresden.de}}\\{\small \it
Institut f\"ur Theoretische Physik, Technische Universit\"at
Dresden,}\\{\small \it D-01062 Dresden, Germany}}
\date{}
\begin{document}

%




\maketitle
\begin{abstract}
 The linear and the nonlinear non-Markovian quantum state diffusion equation
 (NMQSD) are well known tools for the description of certain non-Markovian open
 quantum systems. In this work, we systematically investigate whether the
 normalized linear NMQSD or the nonlinear NMQSD solutions can be generated by
 means of a time-continuous measurement. By considering any conceivable
 measurement scheme in the framework of instruments, we derive a necessary
 criterion for a measurement
 interpretation of both equations. Concrete examples show that the normalized
 linear NMQSD solutions are realizable only in the Markovian limit in general.
 The application of the presented criterion to the nonlinear NMQSD remains an
 open issue.
\end{abstract}
\bigskip
\tableofcontents
\newpage
\section{Introduction}
There is no conceptual problem to assign a pure state to a classical open
system like a Brownian particle. In quantum mechanics however, the entanglement
between the open system and its environment complicates things. Even if the
state of total quantum system is pure, the only state which can be assigned to
an open quantum system is a mixture: the reduced density operator
$\hat\rho_{\rm{red}}(t)$. Obviously, the reduced
density operator can be expressed as a weighted sum over pure states, which is
the starting point for powerful Monte Carlo methods. As a matter of fact, this
so-called \textit{unravelling} is not unique \cite{HJW93}. On the other
hand, a measurement of the environmental degrees of freedom can destroy the
entanglement between the system and its environment. The open system is
therefore in a well defined pure state directly after the measurement.

In order to describe the dynamics of a Markovian open quantum system by means of
Monte Carlo methods, stochastic Schr\"odinger equations are often employed. The
solutions of these evolutions equations for pure states have the remarkable
property that their expectation value coincides with $\hat\rho_{\rm{red}}(t)$.
Yet such stochastic Schr\"odinger equations do not only occur in the context of
Markovian open quantum systems but also as wave function collapse models and
for the description of certain time-continuous measurements
\cite{MSSE,Car93,WM93,SSE_cont_meas,BelBar92}. Moreover, there is a close
relation between time-continuous measurements and the stochastic Schr\"odinger
equations for the description of Markovian open quantum systems: The solutions
of the latter evolution equations can be interpreted as trajectories of
post-measurement states conditioned on a time-continuous measurement. 

In the past fifteen years, stochastic Schr\"odinger equations have been
constructed in order to deal with general non-Markovian open quantum
systems as well. Yet the understanding of these equations in the context of
time-continuous measurements has not been established. Before we come to
this point in more detail, some preliminaries are presented: Firstly, we review
a certain stochastic Schr\"odinger equation approach to non-Markovian open
quantum systems in section \ref{sec_sse_oqs}. Afterwards, the discussion so far
about the relation between these equations and measurement theory is presented
(section \ref{sec_sse_tcm}). In section \ref{sec_tcm}, a general framework for
the description of time-continuous measurements is outlined, which constitutes
the basis for our systematic investigations of the interpretation of this
particular stochastic Schr\"odinger equation in sections \ref{sec_main_part},
\ref{sec_appl} and \ref{sec_nonlin_nmsse}. 

\subsection{Stochastic Schr\"odinger equations and open quantum systems}
\label{sec_sse_oqs}
In many settings, some system of interest, characterized by some Hamiltonian
$\hat H$, is coupled to an environment consisting of finitely or infinitely many
harmonic oscillators, where the coupling is assumed to be linear in the
environmental annihilation and creation operators $\hat a_\lambda$ and
$\hat a_\lambda^\dagger$, respectively \cite{WeissBreu}. Here the structure of
the environment is determined by the oscillator frequencies $\omega_\lambda$
and their distribution. The coupling is characterized by the coupling strengths
$g_\lambda$ which quantify how strongly each environmental mode $\lambda$
couples to the open system via a system coupling operator denoted by $\hat L$.
In the interaction picture with respect to the environmental Hamiltonian 
${\hat H_{env}=\sum_\lambda\omega_\lambda \hat a_\lambda^\dagger\hat
a_\lambda}$, the total Hamiltonian reads
\begin{equation}\label{open_system_model} \hat H_\mathrm{tot}(t)=\hat
H\otimes\mathds{1}+\sum_\lambda \big(\; g^*_\lambda
  \hat L\otimes\hat a^\dagger_\lambda \rme^{\rmi \omega_\lambda t}
  +g_\lambda
  \hat L^\dagger\otimes\hat a_\lambda \rme^{-\rmi \omega_\lambda t}
  \;\big)
\end{equation}
for this class of open quantum systems. In the following, we consider only
scenarios with the initial condition for the total system:
$|\Psi(0)\rangle=|\psi_0\rangle\bigotimes_\lambda|0_\lambda\rangle$.
Here $|0_\lambda\rangle$ denotes the vacuum state with respect to
$\hat a_\lambda$. The environment is hence assumed to be initially at zero
temperature and uncorrelated with the open system. It turns out that the
embedding of the system into its environment is characterized by the
environmental correlation function $\alpha(t-s)$ which equals
\begin{equation}
 \alpha(t-s)=\sum_\lambda |g_\lambda|^2 \rme^{-\rmi \omega_\lambda (t-s)}
\end{equation}
for the above initial condition. Note that this correlation function is
hermitian, which means $\alpha^*(s-t)=\alpha(t-s)$.

If the Born-Markov approximation is applicable,
$\hat\rho_{\mathrm{red}}(t)$ evolves
according to a master equation of the Lindblad
form, where $\hat L$ turns out to be the corresponding Lindbladian
\cite{Car93,WeissBreu,Lind76}. In this paper, an open quantum system is called
Markovian if and only if its reduced density operator obeys a Lindblad master
equation with time-independent Lindbladians. Such a Markovian open system
dynamics can only arise in the microscopic model (\ref{open_system_model}) if
one formally assumes $\alpha(t-s)=\kappa\delta(t-s)$. For large system Hilbert
space dimensions, it is, however, more efficient to employ so-called {\it Monte
Carlo wave function techniques} or {\it quantum trajectories methods} instead of
solving the known operator-valued Lindblad equation numerically
\cite{MSSE,Car93}. These methods are based on stochastic evolution equations for
pure system states as sketched above. 

The dynamics of an open system becomes non-Markovian if the environment is
structured and there is no time-scale separation or if the coupling is strong.
Measures for the non-Markovianity of open system dynamics are currently
discussed \cite{NM_meas}. In this regime, the form of the master equation is in
general unknown. Super-operator techniques are available for determining
$\hat\rho_{\mathrm{red}}(t)$ but often not practical \cite{NakZwan}. Monte
Carlo wave function methods, however, provide an alternative means for coping
with this problem \cite{other_NMSSE1,other_NMSSE2}. In this paper, we
concentrate on an exact generalization of the Markovian quantum state diffusion
equations to the general setting (\ref{open_system_model}) which has been found
by Di\'osi, Gisin and Strunz \cite{DS97, DGS98}. The so-called \textit{linear
non-Markovian quantum state diffusion equation} (linear NMQSD) reads for an
initially zero temperature environment
\begin{equation}\label{lin_nmsse}
 \partial_t |\psi_t([z^*]_0^t)\rangle =
  \big(-\rmi \hat H + z^*_t\hat L - \hat L^\dagger \int_0^t \rmd s\,
  \alpha(t-s)\frac{\delta}{\delta z^*_s}\big)|\psi_t([z^*]_0^t)\rangle,
\end{equation}
where $[z^*]_0^t$ indicates a functional dependence on the drawn path of the
complex Gaussian process $z^*_t$. This stochastic process is completely
characterized by the moments $\mathbb{E}[z_t]=\mathbb{E}[z_tz_s]=0$ and
$\mathbb{E}[z_tz^*_s]=\alpha(t-s)$. By solving (\ref{lin_nmsse}) for a
deterministic initial condition $|\psi_0\rangle$ and many realizations of the
stochastic process $z^*_t$, one can unravel the reduced
density operator of the open system corresponding to 
$\hat\rho_{\mathrm{red}}(0)=|\psi_0\rangle\!\langle \psi_0|$:
\begin{equation}\label{unravel}
 \hat\rho_{\mathrm{red}}(t) = \mathbb{E}
  \big[|\psi_t([z^*]_0^t)\rangle\!\langle \psi_t([z^*]_0^t)|\big].
\end{equation}
Here $\mathbb{E}[\cdot]$ denotes the expectation value over all solutions of
(\ref{lin_nmsse}). When reviewing the derivation of (\ref{lin_nmsse}), one
realizes that for any fixed time $t\geq0$ $|\psi_t([z^*]_0^t)\rangle$ is the
unnormalized relative state of the system corresponding to a coherent
environmental state \cite{DS97}:
Let $|z_\lambda\rangle$ denote the unnormalized Bargmann state with $\langle
z_\lambda|z_\lambda\rangle=\exp(-|z_\lambda|^2)$ corresponding to $\hat
a_\lambda$ and let $\rmd^{2}z_\lambda$ be defined as
$\rmd\mathfrak{Re}(z_\lambda) \rmd\mathfrak{Im}(z_\lambda)$. The state of the
total system at time $t$ can then be represented as
\begin{equation}\label{exp_coherent_states}
 |\Psi(t)\rangle = \int \prod_\lambda\Big(\frac{\rmd^{2}z_\lambda
  \exp(-|z_\lambda|^2)}{\pi}\Big)\quad
 |\psi_t([z^*]_0^t)\rangle \bigotimes_\lambda |z_\lambda\rangle
\end{equation}
if $[z^*]_0^t$ is given by $z^*_\tau=-\rmi\sum_\lambda g_\lambda^* z_\lambda^*
\rme^{\rmi\omega_\lambda \tau}$ where $\tau\in[0,t]$. 

In the special case of a Markovian open quantum system, i.e.
$\alpha(t-s)=\kappa \delta (t-s)$, the linear NMQSD reduces to the well known
linear Markovian quantum state diffusion equation driven by white noise
\cite{DGS98}. It has to be pointed out that the open system dynamics can be
non-Markovian according to the definition given previously while $z^*_t$ is a
Markovian stochastic process. A correlation function for which this is the case
will be discussed in section \ref{sec_appl_dephase}. 

The occurrence of the
functional derivative in (\ref{lin_nmsse}) is a direct
consequence of the driving with coloured Gaussian noise \cite{Budini00}.
Yet if there is a system operator $\hat O(t,s,[z^*]_0^t)$ which removes
the non-locality of (\ref{lin_nmsse}) with respect to the stochastic process,
i.e. 
\begin{equation}\label{ansatz_op}
 \frac{\delta}{\delta z^*_s}|\psi_t([z^*]_0^t)\rangle=
  \hat O(t,s,[z^*]_0^t)|\psi_t([z^*]_0^t)\rangle,
\end{equation}
(\ref{lin_nmsse}, \ref{unravel}) can be used as a Monte Carlo scheme for
calculating $\hat\rho_{\mathrm{red}}(t)$. If such a replacement is feasible,
the linear NMQSD becomes
\begin{equation}\label{lin_nmsse_local}
  \partial_t |\psi_t([z^*]_0^t)\rangle =
  \big(-\rmi \hat H + z^*_t\hat L - \hat L^\dagger \int_0^t \rmd s\,
  \alpha(t-s)\,\hat O(t,s,[z^*]_0^t)\big)\,|\psi_t([z^*]_0^t)\rangle,
\end{equation} 
and one can apply parallel computing in order to solve it for many
process realizations. There are many examples where the replacement
(\ref{ansatz_op}) exists and an exact ansatz operator can be stated. Two of them
will be reviewed in section \ref{sec_appl}. In other cases, approximation
schemes have to be employed in order to calculate $\hat O(t,s,[z^*]_0^t)$
\cite{Ansatz_appr}. As we will see in section \ref{sec_main_part},
(\ref{lin_nmsse_local}) leads to a reversible dynamics of the relative states
$|\psi_t([z^*]_0^t)\rangle$. This does not imply that
$\hat\rho_{\mathrm{red}}(t)$ evolves reversibly, of course. Since the right hand
side of (\ref{lin_nmsse_local}) is not anti-hermitian, the squared norm of a
solution $|\psi_t([z^*]_0^t)\rangle$ is not preserved under the dynamics. Yet
its expectation value remains constant \cite{DGS98}. This observation becomes
important for the following idea.

For estimating the necessary sample size for a good Monte Carlo simulation of
$\hat\rho_{\mathrm{red}}(t)$, it is instructive to average over normalized
states
$|\hat\psi_t([z^*]_0^t)\rangle=|\psi_t([z^*]_0^t)\rangle/||\psi_t([z^*]_0^t)||$
in (\ref{unravel}). If $\nu_0^t(\cdot)$ denotes the Gaussian probability measure
of the paths $[z^*]_0^t$, (\ref{unravel}) can equivalently be written as
\begin{equation}\label{unravel_normal}
 \hat\rho_{\mathrm{red}}(t) = \int
  |\hat\psi_t([z^*]_0^t)\rangle\!\langle \hat\psi_t([z^*]_0^t)|\quad
  ||\psi_t([z^*]_0^t)||^2\nu_0^t(\rmd[z^*]_0^t).
\end{equation}
Here and from now on, we suppress a possible dependence on the
complex-conjugated process realization $[z]_0^t$ in the notation\footnote{The
path probability measure $\nu_0^t$ actually depends on both $[z*]_0^t$ and
$[z]_0^t$.}. According to the observations lined out above, the probability
weight with which $|\hat\psi_t([z^*]_0^t)\rangle$ contributes in
(\ref{unravel_normal}),
i.e. $||\psi_t([z^*]_0^t)||^2\nu_0^t(\rmd[z^*]_0^t)$, changes in the course of
time. Consequently, the necessary sample size for a good estimate of
$\hat\rho_{\mathrm{red}}(t)$ depends on the instant $t$, which makes a
simulation of the dynamics of $\hat\rho_{\mathrm{red}}(t)$ difficult. 

The numerical efficiency can significantly be improved by a Girsanov
transformation, which results in the norm-preserving nonlinear NMQSD
\cite{DGS98}. The latter refers to the evolution equation for
the normalized states
\begin{equation}\label{nonlin_nmsse_sol}
 |\tilde \psi_t([\tilde z^*]_0^t)\rangle := 
  |\hat\psi_t([\tilde z^*]_0^t)\rangle,
\end{equation}
where $\tilde z_t$ denotes the shifted process
\begin{equation}\label{shifted_proc}
 \tilde z_t := z_t+\int_0^t \rmd s\,\alpha(t-s)\,
 \langle \tilde\psi_s([\tilde z^*]_0^s)|
 \hat L|\tilde \psi_s([\tilde z^*]_0^s)\rangle.
\end{equation}
This process has to be regarded as a functional of the coloured noise $z_t$,
which we omit in the notation. In this sense, the reduced density operator is
also unravelled by the nonlinear NMQSD solutions:
\begin{equation}\label{unravel_nonlin}
 \hat\rho_{\mathrm{red}}(t) = \int
  |\tilde\psi_t([\tilde z^*]_0^t)\rangle\!\langle 
    \tilde\psi_t([\tilde z^*]_0^t)|\quad
  \nu_0^t(\rmd[z^*]_0^t).
\end{equation}
Consequently, the Girsanov transformation leads to a stochastic Schr\"odinger
equation whose solution occur with a time-independent weight in the
unravelling.

In the following, we refer to both the normalized solutions
$|\hat\psi_t([z^*]_0^t)\rangle$ of the linear NMQSD and
to the nonlinear NMQSD solutions $|\tilde\psi_t([\tilde z^*]_0^t)\rangle$ as
\textit{(non-Markovian) quantum trajectories}. As stated above, the expectation
values of those objects have a well defined physical meaning. 
This paper is devoted to the question whether a single quantum trajectory is
physically realizable, too. 

\subsection{Stochastic Schr\"odinger equations and time-continuous
measurements}
\label{sec_sse_tcm}
Whether quantum trajectories are subjectively real \cite{Wise96}, i.e. can be
interpreted as trajectories of post-measurement states of a time-continuous
measurement scheme, is a question motivated by both conceptual and practical
interest: The variety of distinct unravellings could possibly be understood as a
consequence of distinct measurement schemes \cite{WV98,WD01}. From an
experimental point of view, a measurement scheme which generates quantum
trajectories would feature a non-demolition property: If applied non-selectively
upon many identical copies of the system, this scheme generates the state of the
unmeasured open quantum system, i.e. $\hat\rho_{\mathrm{red}}(t)$. Certainly,
such a measurement scheme would depend on $\alpha(t-s)$. So one could simulate
the embedding of an open system into differently structured environments.

In the Markovian regime, the interpretation of quantum trajectories is well
established: Quantum state diffusion driven by real (complex) white noise
can be realized by a homodyne (heterodyne) detection of
the environmental photon field \cite{Car93,WM93}. There are many other examples
in which Markovian stochastic Sch\"odinger equations have been postulated in
order to model imprecise time-continuous measurements or the dynamics of the
wave function reduction (e.g. \cite{SSE_cont_meas}). Moreover, Wiseman and
Di\'osi have completely parameterized all diffusive stochastic Schr\"odinger
equations for Markovian open systems and linked those equations with generating
measurement schemes \cite{WD01}. Some of those equations generate a
non-Markovian system state dynamics while the expectation value of those
states obeys a Lindblad master equation. Furthermore, the general form of a
stochastic Schr\"odinger equation driven by independent Wiener
processes has been derived such that it can be interpreted as an evolution
equation for post-measurement states \cite{BG09}. If one averages the
post-measurement states over all measurement records, the resulting {\it a
priori} state evolves according to a master equation of the Lindblad type.

However, the most general class of diffusive stochastic Schr\"odinger equations
which allow a measurement interpretation has not been characterized in the
non-Markovian regime. There are examples where such an interpretation is
indeed feasible \cite{BPP10} and, recently, Yang, Miao and Chen
proposed an interesting approach relating
non-Markovian open quantum systems to time-continuous measurements \cite{YMC11}.
In this context, see also \cite{LD11}. Yet in
particular, the meaning of the
non-Markovian quantum trajectories is a vividly debated open question.

From (\ref{exp_coherent_states}), it becomes clear that for any time $t\geq 0$
the open system can be forced into a pure state proportional to
$|\psi_t([z^*]_0^t)\rangle$ by projectively measuring
in which coherent state $\bigotimes_\lambda|z_\lambda\rangle$ the environment
is. In this context, see also the work of Gambetta and Wiseman \cite{GW02}, who
furthermore show that such a single-shot measurement interpretation leads to
alternative unravellings as well. If, however, those single-shot measurements
are used to constitute a time-continuous measurement scheme performed upon a
single copy of the system, the system dynamics will not be given by a NMQSD
because of the incompatibility of the single-shot measurements. This fact has
motivated an interpretation of NMQSDs as a hidden variable theory \cite{GW03}.
Furthermore, Di\'osi suggested that a certain readout schedule of initially
entangled von Neumann detectors could induce the NMQSD dynamics
\cite{Dios90-08}. In response, Gambetta and Wiseman showed that this scheme does
not preserve the purity of the system states \cite{GW08}. 

In \cite{GW02,GW03,GW08}, Gambetta and Wiseman argue that pure-state quantum
trajectories do not exist for general non-Markovian systems within the framework
of standard quantum mechanics. Their conclusion, however, is based only on a
concrete measurement scheme which fails to generate non-Markovian quantum
trajectories. In particular, the line of argument in \cite{GW02} neglects the
possibility of a measurement scheme which adapts conditioned on previous
outcomes. Therefore, the central aim of this work is to investigate the
realizability of non-Markovian quantum trajectories systematically by
considering any conceivable time-continuous measurement scheme.

\subsection{General description of time-continuous measurements}
\label{sec_tcm}
The general framework for describing quantum measurements was studied by
Davies and Lewis \cite{DL70}, who have introduced the concept of (completely
positive) instruments. In particular, this concept covers von
Neumann measurements \cite{LueNeu} but also indirect and
destructive measurements (e.g. photodetection). A more recent introduction 
into general quantum measurements and instruments can be found in \cite{MW09}
and \cite{Hol01}, respectively. 

Let $\Omega$ be the set of all possible measurement outcomes and let
$\mathcal{F}$ be a $\sigma$-algebra over $\Omega$ such that it contains all
events which can be verified or falsified by a measurement. An event
$F\in\mathcal{F}$ is called verified in an actual measurement if the
measurement outcome $\omega\in\Omega$ is an element of $F$. Then a map
${Y(\cdot)[\cdot]:\;\mathcal{F}\times\mathcal{B}(\mathds{H})\rightarrow  
\mathcal{B}(\mathds{H})}$, where $\mathcal{B}(\mathds{H})$ denotes all bounded
linear operators on the system Hilbert space $\mathds{H}$, is called an
\textit{instrument} if the following conditions are fulfilled: For any
$F\in\mathcal{F}$, $Y(F)[\cdot]$ is an operation, i.e. linear,
trace-decreasing
and completely positive. Besides, the normalization $\tr(Y(\Omega)[\hat
A])=\tr(\hat A)$ is satisfied for any $\hat A$. Finally, $Y(\cdot)[\hat A]$ is
$\sigma$-additive for any $\hat A$.

Given the pre-measurement state $\hat\rho$, the probability for verifying
$F\in\mathcal{F}$ is postulated to be
\begin{equation}\label{postulate_prob}
 P(F|\hat\rho):=\tr(Y(F)[\hat\rho])\equiv\tr\big(\hat\rho\,
  Y^\dagger(F)[\mathds{1}]\big),  
\end{equation}
where $Y^\dagger(F)[\cdot]$ denotes the dual map of $Y(F)[\cdot]$. The
corresponding post-measurement state is then
\begin{equation}\label{postulate_post_meas_state}
 \hat\rho_F:=\frac{Y(F)[\hat\rho]}{\tr(Y(F)[\hat\rho])}.  
\end{equation}
A dilation theorem allows the interpretation of the stochastic dynamics
(\ref{postulate_prob},\ref{postulate_post_meas_state}) as being induced by a
projective measurement upon a measurement apparatus which has been entangled
with the system by some interaction \cite{Ozawa84}. Such an extension is called
a \textit{measuring process} and is not in general unique. 

A general time-continuous measurement can be described either by explicitly
modelling the interplay of general single-shot measurements and short periods of
free evolution (e.g. \cite{BelBar92}) or as measurements that formally follow
the mathematical framework of single-shot measurements (e.g.
\cite{BNL02,BG09}). 
Although the latter description of time-continuous measurements appears to be
rather formal, it is the appropriate means to deal with the issue of this paper.
The advantage of the latter description in comparison to the former is that
one need not know which physical quantity has to be measured after the short
periods of free evolution. 

The chosen approach starts with the continuum limit: The measurement
record is assumed to be a function of time instead of single numbers from the
beginning. So for describing a measurement within the time interval $[0,t]$,
the set of all measurement outcomes $\Omega_0^t$ needs to be an appropriate
space of functions, e.g. all real- or complex-valued right continuous functions
on $[0,t]$ with left limits. We denote a single record by
$[x]_0^t\in\Omega_0^t$, i.e. ${[x]_0^t:=(\tau\mapsto x(\tau), \tau\in[0,t])}$.
Naturally, the measurement record $x(\cdot)$ must be a non-anticipating
stochastic process, i.e. adapted with respect to some filtration
$\{\mathcal{F}_t\}_{t>0}$.

Then a family
of instruments
${\mathcal{Y}=\{Y_0^t(\cdot)[\cdot]:\mathcal{F}_t\times\mathcal{B}(\mathds{H})
\rightarrow \mathcal{B}(\mathds{H});\;t>0\}}$ describes the whole measurement.
In this approach, causality has
to be ensured manually: Let $F_0^s\in\mathcal{F}_s$ be some event which can be
verified or falsified by a measurement up to time $s$. The probability for
verifying $F_0^s$ must not depend on whether or not one measures longer than
necessary, say up to time ${t>s}$, and then throws the additional data away. In
other words, $\mathcal{Y}$ must fulfil
\begin{equation}
  P_0^s(F_0^s|\hat\rho)=\tr(Y_0^s(F_0^s)[\hat\rho])
  =P_0^t(F_0^s\times\Omega_s^t|\hat\rho) 
  =\tr(Y_0^t(F_0^s\times\Omega_s^t)[\hat\rho])
\end{equation}
for any pre-measurement state $\hat\rho$. Here
$F_0^s\times\Omega_s^t$ refers to all elements of $\Omega_0^t$ which lie in
$F_0^s$ if restricted to $[0,s]$. Consequently, $\mathcal{Y}$ can only
represent a time-continuous measurement scheme if the following compatibility
demand is satisfied
\begin{equation}\label{instr_comp}
 \forall 0<s<t,\quad\forall F_0^s\in\mathcal{F}_s:\quad 
  \big(Y_0^s\big)^\dagger(F_0^s)[\mathds{1}]=
  \big(Y_0^t\big)^\dagger(F_0^s\times\Omega_s^t)[\mathds{1}].
\end{equation}
According to a theorem proven by Loubenets, any instrument can be represented
in a particular integral form \cite{Lou01}. As the realizability of certain pure
post-measurement state trajectories is studied in this work, we only state the
representation of instruments corresponding to efficient measurements. In this
context, a measurement is called efficient if a pure pre-measurement state is
turned into a pure post-measurement state as a consequence of the measurement.
So if $\mathcal{Y}$ represents an efficient time-continuous measurement,
there are operator-valued functions $[x]_0^t\mapsto\hat V_s^t([x]_0^t)$, the
\textit{stochastic evolution operators}, and finite scalar measures
${\mu_0^t:\mathcal{F}_t\rightarrow I\subset[0,\infty)}$ for any ${0\leq s<t}$
such that
\begin{equation}\label{instr_rep}
 Y_0^t(F_0^t)[\hat\rho]=\int_{F_0^t}
  \hat V_0^t([x]_0^t)\;\hat\rho\;\big(\hat V_0^t([x]_0^t)\big)^\dagger
  \quad \mu_0^t(\rmd [x]_0^t).
\end{equation}
The probability for finding a measurement record in the vicinity of $[x]_0^t$
reads
\begin{equation}\label{stat_simple}
 P_0^t(\rmd [x]_0^t|\psi)=||\hat V_0^t([x]_0^t)\psi||^2\quad
 \mu_0^t(\rmd[x]_0^t),
\end{equation}
given the initial pre-measurement state $|\psi\rangle$. The corresponding
post-measurement state equals
\begin{equation}\label{post_meas_simple}
 |\psi([x]_0^t)\rangle=\frac{\hat V_0^t([x]_0^t)|\psi\rangle}
      {||\hat V_0^t([x]_0^t)\psi||}.
\end{equation}
By stating conditions on $\hat V_s^t([x]_0^t)$ and $\mu_0^t(\cdot)$ such that
the compatibility condition (\ref{instr_comp}) is satisfied, Barndorff-Nielsen
and Loubenets have presented a general framework for the description of
time-continuous measurements \cite{BNL02}. When applying this formalism to our
problem, we start with determining the ingredients of (\ref{instr_rep}) and then
rederive those conditions on the ingredients from the compatibility demand
(\ref{instr_comp}). In particular, those conditions ensure that the instrument
for an observation within $[0,t]$ can be represented as a composition of
arbitrarily many instruments describing a sequence of measurements up to time
$t$. Namely for any $s\in(0,t)$, one can write
\begin{equation}\label{instr_composition}
 Y_0^t(\rmd[x]_0^s\times \rmd[x]_s^t)[\hat\rho]=
  Y_s^t(\rmd[x]_s^t|[x]_0^s)[Y_0^s(\rmd[x]_0^s)[\hat\rho]],
\end{equation}
where $Y_s^t(\cdot|[x]_0^s)[\cdot]$ denotes an instrument which depends on the
previously measured signal $[x]_0^s$:
\begin{equation}
 Y_s^t(\rmd[x]_s^t|[x]_0^s)[\hat C]:=\hat V_s^t([x]_0^t)\;\hat C\;\big(\hat
    V_s^t([x]_0^t)\big)^\dagger\quad \mu_s^t(\rmd [x]_s^t|[x]_0^s).
\end{equation}
Here the conditional measure ${\mu_s^t(\rmd
[x]_s^t|[x]_0^s)=\mu_0^t(\rmd[x]_0^s\times \rmd[x]_s^t)/\mu_0^s(\rmd [x]_0^s)}$
has been introduced. The possible dependence of $Y_s^t$ on the result $[x]_0^s$
can be seen as a generalization of instrumental processes with independent
increments \cite{InstrProc, Hol01}: The time-continuous measurements described
in \cite{BNL02} are allowed to adapt conditioned on previous measurement
outcomes. 
\section[A necessary criterion]{A necessary criterion for a measurement
interpretation}\label{sec_main_part}

In the following, we consider only NMQSDs whose ansatz operator
is noise-independent\footnote{The proof presented in this paper is more
general: It also holds for ansatz operators which feature an additional linear
functional of the noise. In the context of energy transport in molecules, a
functional Taylor expansion in the noise has recently been applied in order to
approximate the ansatz operator \cite{Ansatz_appr}. The line of argument
presented here is therefore valid for systems for which this Taylor series
terminates after the second term.}, i.e. $\hat
O(t,s,[z^*]_0^t)=\hat O(t,s)$ with $\hat O(s,s)=\hat L$.  In order to study the
transition from the non-Markovian to the Markovian regime, we do not specify
the correlation function $\alpha(t-s)$.

Suppose there is an interpretation of the normalized solution of
(\ref{lin_nmsse_local}) in terms of some time-continuous measurement scheme.
Then there must be a family of instruments $\mathcal{Y}$ such that the
following demands are satisfied:
\begin{enumerate}[i.]
 \item These instruments are compatible with respect to time, i.e. obey
	(\ref{instr_comp}).
 \item Given some fixed pure initial pre-measurement state, the set of all
	post-measurement state trajectories coincides with the set of all
	normalized linear NMQSD solutions corresponding to the given initial
	condition.
 \item If an event $F_0^t\in\mathcal{F}_t$ is associated with a zero
	probability set of normalized linear NMQSD solutions, $F_0^t$
	shall be verified with zero probability in an experiment.
\end{enumerate}
The minimality demand (iii) ensures
that the measurement scheme generates only those normalized NMQSD solutions as
post-measurement state trajectories which actually contribute
to the unravelling of the reduced density operator (\ref{unravel_normal}).
Alternatively, we could have demanded that the {\it a priori}
state of the measurement scheme equals the reduced density operator. Yet such a
stronger demand would blur a result which we anticipate here: Up to a certain
degree, it turns out to be unimportant whether one investigates a possible
measurement interpretation of the linear or of the nonlinear NMQSD. Both issues
can be handled within a single framework - at least in principle. 

The derivation of a necessary criterion for the measurement interpretation of
non-Markovian quantum trajectories consists of three steps: First of all, it
has to be proven which quantity can be assumed to be the measurement signal
w.l.o.g. (section \ref{sec_meas_outcome}).
Secondly, the ingredients of the integral representation (\ref{instr_rep}) are
derived (sections \ref{sec_stoch_evol_op} and \ref{sec_input_meas}). Finally,
the compatibility demand (\ref{instr_comp}) is reformulated in section
\ref{sec_nec_crit}. The resulting criterion can only be a necessary one since we
deal with the problem on the level of instruments without referring to any
experimentally realizable measuring process (cf.\cite{Hol01}).

\subsection{Measurement outcome}\label{sec_meas_outcome}
Given some measurement signal, say $[Z]_0^t$, we must be capable of determining
the corresponding trajectory of post-measurement states up to time $t$ by
assumption. So we assume the existence of a map
$\Phi_t$ from the set of all measurement records $\{[Z]_0^t\}$ to the set of all
normalized linear NMQSD solutions:
$\tau\mapsto|\hat\psi_\tau([z^*]_0^\tau)\rangle$ with $\tau\in[0,t]$ and $|\hat
\psi_0(z^*_0)\rangle =|\psi_0\rangle$. This map is surjective because of
demand (ii). If two measurement signals $[Z_1]_0^t$,
$[Z_2]_0^t$ are associated with the same trajectory of post-measurement
states, the measurement signal can be post-processed such that only
$[Z_1]_0^t$ is displayed as the measurement outcome in both cases.
Consequently, $\Phi_t$ is a bijection w.l.o.g..

Any process realization $[z^*]_0^t$ can be associated with a
normalized solution of the linear NMQSD. Conversely, a given normalized linear
NMQSD solution can be differentiated with respect to time. Due to the
at most linear dependence of the ansatz operator on the noise, the corresponding
process realization $[z^*]_0^t$ can iteratively be determined by comparing that
time-derivative with the known equation of motion. Hence, there is a bijection,
say $\Sigma_t$, from the set of all normalized linear NMQSD solutions within
$[0,t]$ into the set of all process realizations $\{[z^*]_0^t\}$. As a result, 
$\Sigma_t\circ\Phi_t$ provides a one to one correspondence between the
measurement signals $[Z]_0^t$ and the process realizations
$[z^*]_0^t$. For this reason, $[z^*]_0^t$ can be assumed as the measurement
signal w.l.o.g..

\subsection{Stochastic evolution operators}\label{sec_stoch_evol_op}
Let $\hat G_t([z^*]_0^t)$ be the so-called \textit{stochastic propagator} which
maps the initial system state $|\psi_0\rangle$ to the linear NMQSD solution at
time $t$, i.e. $|\psi_t([z^*]_0^t)\rangle=\hat G_t([z^*]_0^t)
|\psi_0\rangle$. It is easy to realize that the ansatz
operator $\hat O(t,s,[z^*]_0^t)$ exists if and only if $\hat
G_t([z^*]_0^t)$ is invertible: If $\hat G_t([z^*]_0^t)$ is non-singular, the
ansatz 
\begin{equation}\label{ansatz_ansatz_op}
 \hat O(t,s,[z^*]_0^t)= 
    \Big(\frac{\delta}{\delta z^*_s}\hat G_t([z^*]_0^t)\Big)\;
    \Big(\hat G_t([z^*]_0^t)\Big)^{-1}  
\end{equation}
certainly satisfies (\ref{ansatz_op}). Conversely, if $\hat O(t,s,[z^*]_0^t)$
exists, one
can integrate the evolution equation for $\hat G_t([z^*]_0^t)$, i.e. effectively
(\ref{lin_nmsse_local}). So $\hat G_t([z^*]_0^t)$ turns out to be a time-ordered
exponential, which is indeed invertible. This observation allows us to define a
\textit{two times propagator} for any $0\leq s \leq t$
\begin{equation}\label{2times_prop}
 \hat A_s^t([z^*]_0^t) := 
     \hat G_t([z^*]_0^t)\,\big( \hat G_s([z^*]_0^s)\big)^{-1}.
\end{equation}
The two times propagator obviously obeys 
$\hat A_s^t([z^*]_0^t)|\psi_s([z^*]_0^s)\rangle = |\psi_t([z^*]_0^t)\rangle$
and satisfies the cocycle condition
\begin{equation}\label{cocycle}
 \hat A_s^t([z^*]_0^t) = \hat A_r^t([z^*]_0^t)\,\hat A_s^r([z^*]_0^r)
\end{equation}
for any $0\leq s\leq r\leq t$. With a glance at postulate
(\ref{post_meas_simple}), we can rewrite
demand (ii) equivalently as $\hat V_0^t([z^*]_0^t)\propto \hat
A_0^t([z^*]_0^t)$. Without loss of generality, the absolute square of the
proportionality factor can be absorbed in the scalar measure
$\mu_0^t(\cdot)$. So up to an irrelevant phase factor, we may identify 
$\hat V_0^t([z^*]_0^t) = \hat A_0^t([z^*]_0^t)$. 

\subsection{Scalar measures}\label{sec_input_meas} 
Recalling that $\nu_0^t(\cdot)$ denotes the path probability measure of the
Gaussian process $[z^*]_0^t$, we explicate the minimality demand as
follows: Any event $F_0^t\in\mathcal{F}_t$ with $\nu_0^t(F_0^t)=0$ is verified
in a measurement up to time $t$ with the probability $P_0^t(F_0^t|\hat\rho)=0$,
independently of the pre-measurement state $\hat\rho$. So demand (iii)
translates to an absolute continuity of the operator-valued measure
$(Y_0^t)^\dagger(\cdot)[\mathds{1}]$ with respect to
$\nu_0^t(\cdot)$:
\begin{equation}\label{abs_cont}
 \forall F_0^t\in\mathcal{F}_t:\quad\nu_0^t(F_0^t)=0
 \quad\Rightarrow\quad \big( Y_0^t\big)^\dagger(F_0^t)[\mathds{1}]=0.
\end{equation}
The finite, positive scalar measure $\mu_0^t$ can be decomposed into a singular
part $\mu^t_\mathrm{s}$ and an absolutely continuous part $\mu^t_\mathrm{a}$
with respect to $\nu_0^t$ by means of Lebesgue's decomposition theorem. 
The singular part $\mu^t_\mathrm{s}$ is concentrated on a set
$C_0^t\in\mathcal{F}_t$ with $\nu_0^t(C_0^t)=0$ and
$\mu^t_\mathrm{s}(\Omega_0^t\setminus C_0^t)=0$. Due to (\ref{abs_cont}), one
finds for the dual instrument
\begin{equation}\label{decomposition}
 \big(Y_0^t\big)^\dagger(C_0^t)[\mathds{1}]=\int_{C_0^t} 
   \big( \hat A_0^t([z^*]_0^t)\big)^\dagger\hat A_0^t([z^*]_0^t)
    \quad \mu^t_\mathrm{s}(\rmd[z^*]_0^t)=0.
\end{equation}
The integrand of (\ref{decomposition}) is a strictly positive definite operator
as the two times propagator is invertible. Therefore,
$\mu^t_\mathrm{s}(C_0^t)=0$ must hold and, consequently,
$\mu^t_\mathrm{s}(\cdot)$ vanishes everywhere. According to the theorem of
Radon-Nikodym, we may conclude that there is
a $\nu_0^t$-a.e. unique positive functional $f_0^t([z^*]_0^t)$ with
$\mu_0^t(\rmd[z^*]_0^t)=f_0^t([z^*]_0^t)\,\nu_0^t(\rmd[z^*]_0^t)$. 

How can these functionals $f_0^t([z^*]_0^t)$ be interpreted? Whereas 
${\hat V_0^t([z^*]_0^t) = \hat A_0^t([z^*]_0^t)}$ ensures that the
post-measurement states depend on the measurement signal in the same way as the
normalized linear NMQSD solutions depend on the process realization,
$f_0^t([z^*]_0^t)$ determines the statistics of the measurements due to
(\ref{stat_simple}). Consequently, the family of functionals $f_0^t([z^*]_0^t)$
can be seen as a means to implement the non-demolition property
(\ref{unravel_normal}). In addition to this, $f_0^t([z^*]_0^t)$ can be regarded
as the Jacobian of a transformation from the actually measurement signal
$[Z]_0^t$ to the process $[z^*]_0^t$. However, not every choice of these
functionals results in a consistent family of instruments, which is the issue of
the next section.

\subsection{A necessary criterion}\label{sec_nec_crit}
According to the previous sections, the normalized linear NMQSD solutions can
only be generated as post-measurement states if the instruments representing
the time-continuous measurement are of the following form
\begin{equation}\label{most_gen_instr}
 Y_0^t(F_0^t)[\hat\rho] = \int_{F_0^t} 
    \hat A_0^t([z^*]_0^t)\,\hat\rho\,\big(\hat A_0^t([z^*]_0^t)\big)^\dagger
    \quad f_0^t([z^*]_0^t)\,\nu_0^t(\rmd[z^*]_0^t).
\end{equation}
Each map (\ref{most_gen_instr}) is an instrument if and only if it is
normalized, i.e.
\begin{equation}\label{normalization}
  \big(Y_0^t\big)^\dagger(\Omega_0^t)[\mathds{1}] = \int_{\Omega_0^t} 
    \big(\hat A_0^t([z^*]_0^t)\big)^\dagger\hat A_0^t([z^*]_0^t)
    \quad f_0^t([z^*]_0^t)\,\nu_0^t(\rmd[z^*]_0^t)=\mathds{1},
\end{equation}
which is a constraint on $f_0^t([z^*]_0^t)$. 
Moreover, the compatibility (\ref{instr_comp}) has to be ensured. Let us fix
$0<s<t$ and some $F_0^s\in\mathcal{F}_s$. Omitting all arguments, we rewrite
the compatibility condition as
\begin{align}\label{comp}\nonumber
 \mathbb{E}_{\nu_0^s}\big[\mathbf{1}_{F_0^s}\,f_0^s\,
		\big(\hat A_0^s\big)^\dagger \hat A_0^s \big]&=
 \mathbb{E}_{\nu_0^t}\big[\mathbf{1}_{F_0^s\times\Omega_s^t}\,f_0^t\,
		\big(\hat A_0^t\big)^\dagger \hat A_0^t \big]\\
 &= \mathbb{E}_{\nu_0^s}\big[\mathbf{1}_{F_0^s}\,
    \mathbb{E}_{\nu_s^t}\big[f_0^t\,
    \big(\hat A_0^t\big)^\dagger\hat A_0^t\big]\big].
\end{align}
Here $\mathbf{1}_A(\cdot)$ refers to the indicator function of the set $A$ and
$\mathbb{E}_{\nu_s^t}[\cdot]$ denotes the conditional expectation value given
some path $[z^*]_0^s$. As (\ref{comp}) holds for any
$F_0^s\in\mathcal{F}_s$, one may conclude
\begin{equation}
 f_0^s\,\big(\hat A_0^s\big)^\dagger \hat A_0^s
  = \mathbb{E}_{\nu_s^t}\big[f_0^t\,\big(\hat A_0^t\big)^\dagger\hat A_0^t\big]
  \qquad\nu_0^s\mathrm{-a.e.}.
 \end{equation}
Because of the cocycle property (\ref{cocycle}) and the two times propagator
being invertible, the last identity may be written as
\begin{equation}\label{compatibility}
   \mathds{1}\,f_0^s([z^*]_0^s)= \int_{\Omega_s^t} 
    \big(\hat A_s^t([z^*]_0^t)\big)^\dagger\hat A_s^t([z^*]_0^t)
    \quad f_0^t([z^*]_0^t)\,\nu_s^t(\rmd[z^*]_s^t|[z^*]_0^s)
    \qquad\nu_0^s\mathrm{-a.e.},
\end{equation}
where $\nu_s^t(\cdot|[z^*]_0^s)$ is the conditional path probability measure of
$[z^*]_0^t$ given a path realization $[z^*]_0^s$. To sum up,
(\ref{most_gen_instr}) represents a time-continuous measurement scheme for every
family of positive functionals $f_0^t([z^*]_0^t)$ which obey the families of
functional integral equations (\ref{normalization},\ref{compatibility}).

Now we tighten demand (iii) and require that the probability (density) to
measure $[z^*]_0^t$ shall coincide with the probability weight of the
corresponding normalized linear NMQSD solution in the unravelling
(\ref{unravel_normal}). This requirement necessarily leads to the choice
$f_0^t([z^*]_0^t)\equiv 1$. Then the normalization (\ref{normalization}) is
fulfilled by construction \cite{DGS98}. In the Markovian regime
$\alpha(t-s)=\kappa\delta(t-s)$, the resulting family of
instruments (\ref{most_gen_instr}) coincides with that one presented in
\cite{BG09}. In this limit, Barchielli and Gregoratti have proven in \cite{BG09}
that the compatibility is fulfilled for any Hamiltonian $\hat H$ and any
coupling operator $\hat L$. 
\newpage
\section{Application of the criterion to concrete systems}\label{sec_appl}
In the following, we check the compatibility demand for $f_0^t([z^*]_0^t)\equiv
1$ in the non-Markovian regime. Therefore, we study two simple models for
which all calculations can be performed analytically without any
approximations.
\subsection{Jaynes-Cummings model}\label{sec_appl_jaynes}
The original Jaynes-Cummings model describes the interaction of a two level atom
with a single electromagnetic mode in the dipole interaction and rotating wave
approximation \cite{JC63}. Here we consider a generalized Jaynes-Cummings model
where there is a coupling of the atom to arbitrarily many environmental modes.
In particular, this model has been studied by means of NMQSDs \cite{DGS98}. For
this problem, the system Hamiltonian reads $\hat H=\omega/2\,\hat\sigma_z$,
where $\omega$ is the atomic transition frequency and $\hat \sigma_z$ the Pauli
spin matrix of the $z$-direction. The coupling operator is given by the lowering
operator $\hat L = \hat\sigma_-$. Then the ansatz operator also turns out to be
noise independent: ${\hat O(t,s,[z^*]_0^t)=f(t,s)\,\hat\sigma_-}$. Here $f(t,s)$
is a function obeying $f(s,s)=1$ and
\begin{equation}
 \partial_t f(t,s)=\big(\rmi \omega+F(t) \big)f(t,s),
\end{equation}
where $F(t):=\int_0^t \rmd s\,\alpha(t-s)\,f(t,s)$. The corresponding linear
NMQSD equals
\begin{equation}\label{lin_nmsse_jayn_cum}
 \partial_t
|\psi_t([z^*]_0^t)\rangle=\big(-\rmi\frac{\omega}{2}\,\hat\sigma_z
      + z_t^* \,\hat\sigma_--F(t)\, \hat\sigma_+ \hat \sigma_-\big)\,
      |\psi_t([z^*]_0^t)\rangle.
\end{equation}
Due to the noise-independence of $\hat O(t,s,[z^*]_0^t)$, the master equation
can be determined from (\ref{lin_nmsse_jayn_cum}) \cite{StrYu04}: 
\begin{align}
  \partial_t \hat\rho_{\rm{red}}(t) =
  &-\rmi\,\big[\,\frac{\omega}{2}\hat\sigma_z+\mathfrak{Im}(F(t)) 
    \hat\sigma_+\hat\sigma_-,\,\hat\rho_{\rm{red}}(t)\big]\\\nonumber
  &+\mathfrak{Re}(F(t))
\,\big(\big[\hat\sigma_-\,\hat\rho_{\rm{red}}(t),\,\hat\sigma_+\big] +
  \big[\hat\sigma_-,\,\hat\rho_{\rm{red}}(t)\,\hat\sigma_+\big]\big).
\end{align}
Consequently, the evolution equation for $\rho_{\rm{red}}(t)$ is of the
Lindblad form but with a time-dependent Lindbladian in general. Only for
$\alpha(t-s)=\kappa \delta(t-s)$, the Lindbladian becomes time-independent
because one has then $F(t)\equiv\kappa/2$. For a non-Markovian open quantum
system, the decay rate $\mathfrak{Re}(F(t))$ can become negative, which does
not, however, cause any problems in the NMQSD approach. In this context, see
also \cite{other_NMSSE2}.

One easily finds the two
times propagator for the Jaynes-Cummings model, namely
\begin{equation}
  \hat A_s^t([z^*]_0^t) =
 \left(\begin{array}{cc}
   \rme^{-\rmi\frac{\omega}{2}(t-s) - \int_s^t \rmd\tau\,F(\tau)} &0\\
  \rme^{+\rmi\frac{\omega}{2}(s+t)} \int_s^t \rmd\tau\,
  z^*_\tau\,\rme^{- \rmi \omega \tau
    -\int_s^\tau \rmd\tau'\, F(\tau') }
  &\rme^{+\rmi\frac{\omega}{2}(t-s)}\\ 
 \end{array}\right),
\end{equation}
where $(1,0)^\mathrm{T}$ and $(0,1)^\mathrm{T}$ represent the eigenvectors
of $\hat\sigma_z$ corresponding to the eigenvalues $+1$ and $-1$, respectively.
Remarkably, the two times propagator depends only on the noise realization
within $[s,t]$. This property has been identified by Barndorff-Nielsen and
Loubenets as being one of two conditions which lead to an instrumental
process with independent increments if they are simultaneously fulfilled - given
that the measurement scheme is consistent, of course \cite{BNL02}.
The integrand of the compatibility demand turns out to be of the form
\begin{equation}\label{comp_integrand_jayn}
 \big(\hat A_s^t([z^*]_0^t)\big)^\dagger\hat A_s^t([z^*]_0^t) =
 \left(\begin{array}{cc}
   h_s^t([z^*]_s^t) &  j_s^t([z^*]_s^t)\\
  \big(j_s^t([z^*]_s^t)\big)^*
  &1\\ 
 \end{array}\right).
\end{equation}
The functionals $h_s^t(\cdot)$ and $j_s^t(\cdot)$ are stated in the appendix.
According to the previous section, the following conditions must be
satisfied if the normalized linear NMQSD solutions allow a measurement
interpretation: $\mathbb{E}_{\nu_s^t}[h_s^t]=1$ and
$\mathbb{E}_{\nu_s^t}[j_s^t]=0$ for any given realization $[z^*]_0^s$. We have
explicated both conditions by discretizing the occurring path integrals and
taking the left boundary points of the tiny time intervals as intermediate
points. The resulting high-dimensional Gaussian integrals can be evaluated
analytically. Having performed the continuum limit again, the condition
$\mathbb{E}_{\nu_s^t}[j_s^t]=0$ turns into the following integral equation for
the correlation function
\begin{equation}\label{comp_jaynes_cumm}
 \int_s^t d\tau\,\alpha(\tau-u)\,
    \exp\big(\,\rmi\omega \tau-\int_s^\tau d\tau'\,F^*(\tau')\,\big)=0
\end{equation}
for any $0<s<t$ and any $u\in[0,s)$. This integral equation is certainly
satisfied by the correlation function of a Markovian open quantum system, i.e.
$\alpha(t-s)=\kappa\delta(t-s)$. For this special case, an easy calculation
verifies $\mathbb{E}_{\nu_s^t}[h_s^t]=1$. So the normalized solutions of
(\ref{lin_nmsse_jayn_cum}) allow a measurement interpretation in the Markovian
regime as expected (cf. \cite{Car93,WM93,BG09}). If, however, we restrict
ourselves to functions $\alpha(\cdot)$ which are Lebesgue integrable,
the fundamental theorem of calculus tells us that the only solution of
(\ref{comp_jaynes_cumm}) is $\alpha(\tau)=0$ Lebesgue-a.e.. 
In this sense, the example shows that the normalized linear NMQSD solutions
generically allow a measurement interpretation only in the Markovian regime.
\subsection{Dephasing interaction}\label{sec_appl_dephase}
Let us consider one of the simplest open quantum systems in order to
analyse whether there are certain limits in which a measurement interpretation
is approximately possible. A qubit, described by 
$\hat H = \omega/2\,\hat\sigma_z$, shall be coupled to the bosonic environment
in a purely dephasing way: $\hat L =r/\kappa\,\hat\sigma_z\propto\hat H$. Here
$\kappa$ denotes some time scale and the dimensionless parameter $r$ controls
the interaction strength. It is easy to show that the ansatz operator is
noise-independent for this setting: $\hat O(t,s,[z^*]_0^t)=\hat L$ \cite{DGS98}.
One then obtains
\begin{equation}
  \partial_t |\psi_t([z^*]_0^t)\rangle =
  \big(-\rmi\frac{\omega}{2}\,\hat\sigma_z + z^*_t\frac{r}{\kappa}\sigma_z
   -\frac{r^2}{\kappa^2}\int_0^t \rmd s\,
  \alpha(t-s)\big)|\psi_t([z^*]_0^t)\rangle
\end{equation}
for the linear NMQSD, which can directly be integrated. Similarly to the
previous section, the master equation can be stated and is of the Lindblad form
with an in general time-dependent Lindbladian:
\begin{align}
   \partial_t \hat\rho_{\rm{red}}(t) =
 -\rmi\,\big[\,\frac{\omega}{2}\hat\sigma_z,\,\hat\rho_{\rm{red}}(t)\big]
  +\frac{r^2}{\kappa^2}\mathfrak{Re}(K(t))
\,\big(\big[\hat\sigma_z\,\hat\rho_{\rm{red}}(t),\,\hat\sigma_z\big] +
  \big[\hat\sigma_z,\,\hat\rho_{\rm{red}}(t)\,\hat\sigma_z\big]\big),
\end{align}
where $K(t):=\int_0^t\rmd s\,\alpha(t-s)$.
One finds again that the
two times propagator depends only on the noise realization within $[s,t]$:
\begin{equation}\label{2times_prop_deph}
 \hat A_s^t([z^*]_s^t)=\exp\big(\,-\rmi\frac{\omega}{2}(t-s)\,\hat\sigma_z
	-\frac{r^2}{\kappa^2} \Theta(t,s) 
	+\frac{r}{\kappa}\int_s^t\rmd\tau\,z^*_\tau\,\hat\sigma_z\,\big),
\end{equation}
where $\Theta(t,s):=\int_s^t\rmd\tau'\int_0^{\tau'}\rmd\tau'' 	   
\,\alpha(\tau'-\tau'')$. The discretized path integral of (\ref{compatibility}) 
with $f_0^t\equiv 1$ can be evaluated analytically. The resulting
effective compatibility demand reads
\begin{equation}\label{comp_dephasing}
 \forall0<s<t,\quad \forall u\in[0,s):\quad \int_s^t d\tau\,\alpha(\tau-u)=0
\end{equation}
in the continuum limit. So again a measurement interpretation is exclusively
possible in the Markovian regime.

It is nevertheless instructive to calculate the right hand side of the
compatibility demand for the correlation function of an Ornstein-Uhlenbeck
process,
\begin{equation}
 \alpha(t-s)=\frac{\kappa\Gamma}{2}\,\rme^{-\Gamma|t-s|},
\end{equation}
which tends to $\kappa\,\delta(t-s)$ as $\Gamma\rightarrow\infty$. If one
divides
$(s,t]$ into $N$ tiny intervals of the length $\Delta t=(t-s)/N$ and defines
$z^*_j:=z^*_{s+j\Delta t}$, $j\in\{0,...,N\}$, the conditional path probability
measure can be expressed as
\begin{align}
  \nu_s^t(\rmd[z^*]_s^t|[z^*]_0^s)&=\lim\limits_{N\rightarrow\infty}
    \prod\limits_{j=1}^N\Big(\frac{\rmd^{2}z_j}{\pi \kappa\Gamma^2 
    \Delta t}\Big)\;
    \exp\Big(-\frac{\Delta t}{\kappa\Gamma^2}
\,\sum\limits_{j=1}^N\Big|\frac{z_j-z_{j-1}}{\Delta t}
    +\Gamma z_{j-1}\Big|^2\,\Big)\\\nonumber
  &=\nu_s^t(\rmd[z^*]_s^t|z^*_s).
\end{align}
With respect to this measure, the right hand side of the compatibility demand,
say $\hat C_{RHS}(s,t,[z^*]_0^s)$, turns out to be a path integral with a
quadratic action. Such an expression can be evaluated by employing standard
techniques of path integration (e.g. \cite{Schu81}), which leads to
\begin{align}\label{comp_dephasing_OU}
  \hat C_{RHS}(s,t,[z^*]_0^s)=
   \exp\{&-\frac{1}{ \kappa\Gamma} [r^2(e^{-\Gamma t}-
   e^{-\Gamma s})\\\nonumber&-
2r(\hat\sigma_z\,\mathfrak{Re}(z_s)-r)(1-e^{-\Gamma(t-s)})-\frac{r^2}{2}
(1-e^{-2\Gamma(t-s)})]\}.
\end{align}
As expected for a Markov process like the Ornstein-Uhlenbeck process, this
expression depends exclusively on (the real part of) $z^*_s$ and not on the
whole path $[z^*]_0^s$. The compatibility is clearly fulfilled in the Markovian
limit: 
\begin{equation}
 \lim\limits_{\Gamma\rightarrow\infty}\hat C_{RHS}(s,t,z^*_s)=\mathds{1}.
\end{equation}
Trivially, the compatibility holds in the limit of vanishing interaction
$r\rightarrow0$. The system dynamics then becomes unitary and independent of
the measurement signal $[z^*]_0^t$, which means that the dynamics is induced by
a trivial measurement. Moreover, the compatibility can be achieved approximately
either in the weak coupling and short correlation time limit $\kappa\Gamma\gg1$,
which is the physical Markov limit in fact, or in the short-time limit
$(t-s)\ll1/\Gamma$ and $(t-s)\ll \kappa$. In the latter scenario, the
correlation function is approximately constant on the considered time scales,
$\alpha(t-s)\approx\kappa\Gamma/2$, which results in linear dependence of
$z_t$ on $z^*_s$ almost surely, namely $z_t\approx z^*_s\approx0$. By
inspecting the two times propagator (\ref{2times_prop_deph}), one can easily
show that the system dynamics turns out to be effectively unitary: The
considered times are much smaller than the time scale of the interaction.
Therefore, the system may be regarded as being decoupled from its environment. 
Finally by inspecting (\ref{comp_dephasing_OU}), one realizes that there is
no meaningful long-time limit such as $(t-s)\gg1/\Gamma$ which approximately
leads to compatibility. 

For a fair discussion of these results, one has to
pay respect to the involved time scales. Since the measurement is described by a
family of instruments $\mathcal{Y}$, the continuum limit is anticipated.
Consequently, the time span between two consecutive measurements, which
constitute the time-continuous measurement scheme, is arbitrarily short and
thus the shortest involved time scale - unless the open system is Markovian. In
the latter case, the time span between two consecutive measurements is
comparable with the time scale of the environmental dynamics. 

\section{On the nonlinear NMQSD}\label{sec_nonlin_nmsse}
In this section, we show that the necessary criterion
(\ref{most_gen_instr},\ref{normalization},\ref{compatibility}) can also be used
for investigating a possible measurement interpretation of the nonlinear NMQSD
in principle.

Due to the linear dependence of the ansatz
operator on the noise and the shift relation (\ref{shifted_proc}), there is a
bijection between the process realizations $[z^*]_0^t$ and solutions of the
nonlinear NMQSD up to time $t$. So $[z^*]_0^t$ can be assumed as the measurement
signal w.l.o.g.. Moreover, (\ref{nonlin_nmsse_sol}) indicates that
$|\tilde\psi_t([\tilde z^*]_0^t)\rangle$ depends on the realization $[\tilde
z^*]_0^t$ in the same way as $|\hat\psi_t([z^*]_0^t)\rangle$ depends on
$[z^*]_0^t$. Therefore, we may conclude $\hat V_0^t(\cdot) = \hat A_0^t(\cdot)$
with a glance at (\ref{post_meas_simple}). So the normalized linear NMQSD
solutions can be turned into nonlinear NMQSD solutions by appropriately
weighting the driving process realizations $[z^*]_0^t$. When reviewing the
derivation of the nonlinear NMQSD \cite{DGS98}, one realizes that the path
probability measure of the shifted process $\tilde z^*_t$ is absolutely
continuous with respect to the path probability measure of $z^*_t$, i.e.
$\nu_0^t(\cdot)$. Consequently, the minimality demand (iii) implies
$\mu_0^t(\rmd[z^*]_0^t)=f_0^t([z^*]_0^t)\,\nu_0^t(\rmd[z^*]_0^t)$ again.
As a result, any family of instruments which might generate the nonlinear NMQSD
solutions as trajectories of post-measurement states must be of the form
(\ref{most_gen_instr}).

So if the nonlinear NMQSD can be interpreted in terms of a time-continuous
measurement scheme, there must be a family of functionals $f_0^t([z^*]_0^t)$
such that (\ref{normalization}) and (\ref{compatibility}) hold and that the pure
post-measurement state trajectories occur with the same probability (density) as
they appear as nonlinear NMQSD solutions.
\section{Conclusions} 
In this paper, we have considered NMQSDs whose ansatz operators depend at most
linearly on the stochastic process $z^*_t$. For this class of open quantum
systems, we have derived a necessary criterion for the measurement
interpretation
of non-Markovian quantum trajectories. By employing a representation theorem
for instruments, any time-continuous measurement scheme which is conceivable in
the framework of standard quantum mechanics has been taken into account. 
In particular, this criterion covers both ``direct'' and ``indirect''
measurements. Here the attribute ``direct'' refers to general measurements
which are implemented by \textit{some} measurement apparatus in the sense of the
dilation theorem (cf. \cite{Ozawa84}). In contrast to this, ``indirect''
measurements bear on 
settings where the \textit{given} harmonic environment acts as a quantum probe
(e.g. heterodyne detection). By tracing out the environmental degrees of
freedom, any instrument on the bipartite Hilbert space of the ``indirect''
measurement setting reduces to an instrument acting on the system Hilbert space
as long as the initial state is not entangled. Hence, the above formalism also
applies in this case. 

A word of caution is in order here: The proof presented
in this paper is based on the assumption that any measurement can be described
by a completely positive instrument. Yet an entangled initial state of the total
system could lead to indirect measurements upon the system which cannot be
described by complete positive maps (cf. \cite{Pech04}).

We have especially concentrated on the interpretation of the linear NMQSD: For
a suitable choice of $f_0^t$, the family of instruments (\ref{most_gen_instr})
generates the normalized solutions of the linear NMQSD. In the Markovian regime,
the resulting measurement scheme coincides with the family of instruments
presented in \cite{BG09}, where the compatibility of these instruments is shown
for any finite dimensional system. 

By investigating examples, we have shown that in general the normalized
solutions $|\hat\psi_t([z^*]_0^t)\rangle$ of the linear NMQSD allow a
measurement interpretation only if $\alpha(t-s)=\kappa\delta(t-s)$. The
compatibility of these measurements is
violated in the non-Markovian regime. As our analysis is performed on the level
of instruments without considering a concrete measuring processes, a descriptive
explanation for this failure cannot be given. We may only conclude that in
general causality is violated by any measurement scheme which might generate
the normalized linear NMQSD solutions. At least, one can state
certain limits in which the compatibility is approximately satisfied.

Up to now, it remains an open question whether the nonlinear NMQSD solutions
allow a measurement interpretation in the non-Markovian regime. We hope that
(\ref{most_gen_instr},\ref{normalization},\ref{compatibility}) can explicitly be
applied to some system in order to investigate this question systematically.
Our investigation moreover shows a mathematical peculiarity of the unravelling
(\ref{unravel_normal}) in comparison to (\ref{unravel_nonlin}) in the
non-Markovian regime: In section \ref{sec_appl}, we have effectively proven that
$||\psi_t([z^*]_0^t)||^2$ is not in general a $(\mathcal{F}_t)$-martingale if
$\alpha(t-s)\neq\kappa\delta(t-s)$. This implies that the family of probability
measures $\{\int_\cdot ||\psi_t([z^*]_0^t)||^2\nu_0^t(\rmd [z^*]_0^t), t>0\}$
is not consistent. Consequently, one can numerically perform a Monte Carlo
simulation of $\hat\rho_{\mathrm{red}}(t)$ based on the normalized linear NMQSD
solutions $|\hat\psi_t([z^*]_0^t)\rangle$ but there is no overall path
probability measure for the driving process $z^*_t$ in this case. The Girsanov
transformation, however, shifts this inconsistency of the measures into the
dynamics of the nonlinear NMQSD solutions. The effectively driving process
$z^*_t$ is then Gaussian and one can show that Kolmogorovs extension theorem is
applicable for any $\alpha(t-s)$. 
This difference in the unravellings does not occur for the Markovian quantum
state diffusion as $||\psi_t([z^*]_0^t)||^2$ turns out to be a
$(\mathcal{F}_t)$-martingale (cf. \cite{BG09}).

It would be interesting to invert the question discussed in this paper. Namely,
is there a family of functionals $f_0^t$ such that the functional integral
equations (\ref{normalization},\ref{compatibility}) are satisfied and the {\it a
priori} state $Y_0^t(\Omega_0^t)[|\psi_0\rangle\!\langle\psi_0|]$ coincides with
$\hat\rho_{\rm{red}}(t)$?

Finally, it is important to notice that the proof of
(\ref{most_gen_instr},\ref{normalization},\ref{compatibility}) mainly bases on
measure-theoretical arguments. Therefore, the criterion
(\ref{normalization},\ref{compatibility}) can be applied to a broader class of
stochastic Schr\"odinger equations, which is defined by the following
two demands:
First, any of these stochastic Schr\"odinger equations is either linear or can
be
formulated in a linear version. This linearity implies that the corresponding
stochastic propagator is invertible. Second, this linear equation must
be
linearly driven by some stochastic process.
For any of these stochastic Schr\"odinger equations,
(\ref{most_gen_instr},\ref{normalization},\ref{compatibility})
provides a tool to investigate whether the randomness of its solutions possibly
originates from a time-continuous measurement.
\begin{center}
 \bf Acknowledgement
\end{center}

We would like to thank Elena R. Loubenets for the enlightening
discussion about the quantum stochastic representation of time-continuous
measurements and Philip M. R. Brydon for his helpful comments on
the manuscript. S.K. thanks Benjamin Trendelkamp-Schroer for their inspiring
discussions.

\appendix
\setcounter{section}{1}
\section*{Appendix}
For the Jaynes-Cummings model, the operator-valued integrand of
(\ref{compatibility}) contains the following two functionals
\begin{align}
   h_s^t([z^*]_s^t) &:= \exp\big[- 2\int_s^t
    \rmd\tau\,\mathfrak{Re}(F(\tau))\,\big]\quad+\quad 
    \int_s^t \rmd\tau_1\int_s^t\rmd\tau_2\,z_{\tau_1}\,z^*_{\tau_2}
    \quad\times\\\nonumber
    &\quad\quad\times\quad
     \exp\big[\,\rmi\omega(\tau_1-\tau_2)
     -\int_s^{\tau_1} \rmd\tau_1'\,F^*(\tau_1')
    -\int_s^{\tau_2} \rmd\tau_2'\, F(\tau_2')\,\big],\\
    j_s^t([z^*]_s^t) &:= \int_s^t \rmd\tau\, z_\tau\,
    \exp\big[ \rmi \omega (\tau-s)
    -\int_s^\tau \rmd\tau'\, F^*(\tau')\,\big].
\end{align}

\end{document}